\documentclass[12pt,aip,superscriptaddress,preprint]{revtex4-1}
\usepackage{color}
\usepackage{graphicx,amsmath,amsfonts,amssymb}
\usepackage{tikz,psfrag,graphicx}
\newcommand{\eps}{\varepsilon}
\renewcommand{\vec}[1]{\mathbf{#1}}

\begin{document}

\title{Light scattering by dielectric bodies in the Born approximation}
\author{A. Bereza}
\affiliation{Institute of Automation and Electrometry, Russian Academy of Sciences, Siberian Branch, 1 Koptjug Ave. Novosibirsk 630090, Russia}
\affiliation{Novosibirsk State University, 2 Pirogov Str., Novosibirsk 630090, Russia}
\author{A. Nemykin}
\affiliation{Institute of Automation and Electrometry,  Russian Academy of Sciences, Siberian Branch,  1 Koptjug Ave. Novosibirsk 630090, Russia}
\affiliation{Novosibirsk State University, 2 Pirogov Str., Novosibirsk 630090, Russia}
\author{S. Perminov}
\affiliation{Rzhanov Institute of Semiconductor Physics,  Russian Academy of Sciences, Siberian Branch, 13 Lavrent'yev Ave., Novosibirsk 630090, Russia}
\author{L. Frumin}
\affiliation{Institute of Automation and Electrometry,  Russian Academy of Sciences, Siberian Branch, 1 Koptjug Ave. Novosibirsk 630090, Russia}
\affiliation{Novosibirsk State University, 2 Pirogov Str., Novosibirsk 630090, Russia}
\author{D. Shapiro}
\affiliation{Institute of Automation and Electrometry,  Russian Academy of Sciences, Siberian Branch, 1 Koptjug Ave. Novosibirsk 630090, Russia}
\affiliation{Novosibirsk State University, 2 Pirogov Str., Novosibirsk 630090, Russia}

\begin{abstract}
    Light scattering is one of the most important elementary processes in near-field optics. We build up the Born series for scattering by dielectric bodies with step boundaries. The Green function for a 2-dimensional homogeneous dielectric cylinder is obtained.  As an example, the formulas are derived for scattered field of two parallel cylinders. The polar diagram is shown to agree with numerical calculation by the known methods of discrete dipoles and boundary elements.
    \pacs{03.50.De,42.25.Fx}
\end{abstract}


\maketitle

\section{Introduction}

In the past decades a substantial progress has been achieved in
nano-optics \cite{Novotny_book,Girard05,SK07}. However, a
significant methodological deficiency still persists even for
basic problems, like scattering by nano-sized bodies. Unlike
"macroscopic" optics, where transverse waves (for instance, plane
or spherical) are very useful to study, say, diffraction and
interference, at sub-wavelength region the treatment of these
phenomena becomes much more complicated. The reason is evanescent waves
near a boundary of illuminated objects. Such wave usually can be
neglected while studying optical processes with large scatterers,
but nano-optics is not the case. Strong coupling via evanescent
wave is the key feature, which most practical nanophotonics
tasks focus on. They include light energy concentration within
few-nanometer range \cite{Stockman11}; high-efficiency broad-band
solar cells \cite{oeCD13}; light-induced forces at
nano-scale \cite{Perminov08,Shapiro16};
surfaces-enhanced Raman spectroscopy \cite{arpcWvD07};
the tomographic reconstruction of a nano-structure. \cite{Yamaoki:16}

Only few problems allow analytical solution in photonics. Along with the classical papers devoted to one cylinder, \cite{pmR18,cjpW55} the scattering from two circular cylinders \cite{jV10} and two perfectly
conducting spheres \cite{Mazets00} can be found in the quasi-static limit using bipolar coordinates; a perfectly conducting cylinder near a surface was considered using expansion in the series of cylindrical
waves.\cite{Borghi96} In any more complicated cases, numerical or semi-analytical methods become the only capable to calculate electromagnetic fields in both near and far regions, for instance, in a system of
several cylinders or in their periodic chain.\cite{schaudt1991exact,Belan:15,Lee2016119}

Analytical approximations are very useful for understanding the scattering properties of a structure, at least for testing the numerical methods. There is a universal method to derive the formulas based on the Born
approximation. It consists in taking the incident field in place of the total field at each point inside the scattering potential. If the scatterer is not sufficiently weak, the next approximations are exploited.
There are several recent optical researches devoted to high-order terms of the approximation. In optical diffusion tomography the high orders are necessary for solving the nonlinear inverse problem.
\cite{panasyuk2006nonlinear} The second-order approximation is needed for numerical reconstruction of a shallow buried object by the scattered amplitude. \cite{Salucci:14} The resonant-state expansion approximation
uses the second-order terms to find eigen frequencies in an optical fiber waveguide. \cite{PhysRevA.93.023835} However, the traditional Born series is not applicable in a system of dielectric bodies with step
edges, as not satisfying the boundary conditions.

The main goal of the present work is to construct modified Born approximation for a set of dielectric bodies. The integral relations are derived and the series for two dielectric cylinders is obtained. We manage to
account for the first cylinder exactly by means of the special Green function for a cylindrical dielectric, that intrinsically include multiple scattering processes with this cylinder. Thus, another aim of our
paper is to derive that special Green function.

The Born series is constructed in Sec.~\ref{sec:Born}.  The scattering by two cylinders, considered in Sec.~\ref{sec:pair}, illustrates the application of developed approach. The obtained formulas are in agreement
with numerical calculation using surface integral equations and discrete dipole approximation. The Green function is derived in Appendix~\ref{sec:Green}: the expressions for the source point inside and outside the dielectric are given for both cases of $p$- and $s$-wave. The boundary element method has already been discussed in previous works devoted to the scattering by cylinders on a dielectric
substrate.\cite{olBFPS11,eplBPFS12,joptFNPS13} The formulas for two-dimensional discrete dipole method are derived in Appendix \ref{sec:DDA}.

\section{Born series}\label{sec:Born}

The Helmholtz equations for magnetic field inside and outside the dielectric (denoted by subscripts ${in}$ and ${out}$, correspondingly) are
\begin{eqnarray}\label{Helmholts-1}
(\bigtriangleup+k_1^2)\mathcal{H}_{in}(\vec{r})=0\;,\quad
(\bigtriangleup+k_0^2)\mathcal{H}_{out}(\vec{r})=0,
\end{eqnarray}
where $\bigtriangleup$ is 2-dimensional Laplace operator with respect to $x$ and $y$ variables, wavenumbers  $k_0=\omega/c$ and $k_1=\sqrt{\eps}\omega/c$, $c$ is the speed of
light, $\omega$ is the frequency, $\eps$ is the dielectric permittivity. The field $\mathcal{H}_{out}$ in free space is slightly changed due to a weak perturbation, which is small enough (i.e. $k_1a \ll 1$, where
$a$ is its size) and/or low-polarizable ($|\eps-1|\ll1$). The internal field $\mathcal{H}_{in}$ can be quite different. The Green function obeys the equation
\begin{equation}\label{Helmholtz-equation}
(\bigtriangleup + k^2) G = \delta (\mathbf{r} -
\mathbf{r}').
\end{equation}
Here ${k}$ is the wavenumber in corresponding region: $k=k_0$ or $k=k_1$.

\begin{figure}\centering
    \begin{tikzpicture}
    \filldraw[fill=green!20!white] (0,0) ellipse (40 pt and 20 pt);
    \draw (0,0) node {\large$\mathcal{D}^+$};
    \draw (0,-1.2) node {\large$\gamma$};
    \draw (0,-2.7) node {$\large\Gamma$};
    \draw (2,-1.7) node {$\large\mathcal{D}^-$};
    \draw[dashed,very thick,red] (0,0) circle (3);
    \draw[->,very thick,brown] (1.4,0)--(0.9,0);
    \draw[->,very thick,brown] (3,0)--(3.5,0);
    \end{tikzpicture}
    \caption{The domains of integration $\mathcal{D}^+$ and $\mathcal{D}^-$ for Eq. (\ref{Green-integral2D}). The boundary of $\mathcal{D}^-$ consists of $\gamma=\partial\mathcal{D}^+$ (solid line) and external infinitely remote  contour $\Gamma$ (dashed line). The arrows indicate its external normals $\vec{n}$ to $\partial\mathcal{D}^-$.}\label{fig:infty}
\end{figure}
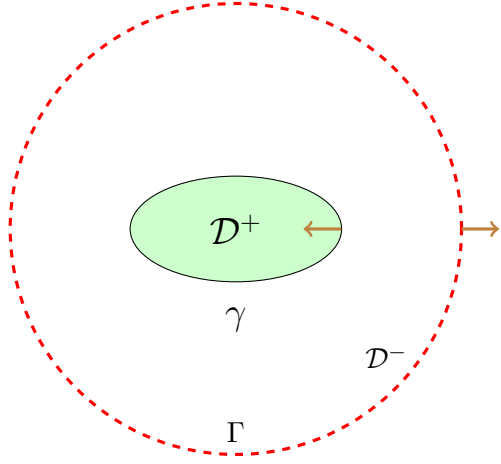

We use Eq. (\ref{Helmholts-1}), (\ref{Helmholtz-equation}) to derive the relations between field amplitudes at the boundary:
\begin{eqnarray}\label{Green-integral2D}
\mathcal{H}_{in}(\vec{r})=\int_{\mathcal{D}^+}\Big[\mathcal{H}_{in}(\vec{r}')\bigtriangleup G_p(\vec{r},\vec{r'})
-G_p(\vec{r},\vec{r'})\bigtriangleup\mathcal{H}_{in}(\vec{r}')\Big]dS'\;,\nonumber\\
\mathcal{H}_{out}(\vec{r})=\int_{\mathcal{D}^-}\Big[\mathcal{H}_{out}(\vec{r}')\bigtriangleup G(\vec{r},\vec{r'}) -
G(\vec{r},\vec{r'})\bigtriangleup\mathcal{H}_{out}(\vec{r}')\Big]dS'\;,
\end{eqnarray}
were $dS'$ is the element of integration over dielectric, $\mathcal{D}^+$, or free space, $\mathcal{D}^-$, domains, Fig.~\ref{fig:infty}. The Green function $G(\vec{r},\vec{r'})$ describes free space, $G_p(\vec{r},\vec{r'})$ is similar function that corresponds to  dielectric of permittivity $\eps$. The Green function of free space is the solution of Eq.
(\ref{Helmholtz-equation}) with $k=k_0$ and can be written as
\begin{eqnarray}
G(\vec{r},\vec{r'}) =\frac1{4i}H_0^{(1)}(k|\vec{r}-\vec{r}'|)= \frac1{4i}\sum_{m=-\infty}^{\infty}e^{im(\varphi-\varphi')}\times
\begin{cases}
H^{(1)}_m(kr\phantom{'})J_m(kr'), & r\phantom{'}>r',\\
H^{(1)}_m(kr')J_m(kr\phantom{'}), & r'>r\phantom{'},\label{free-space}
\end{cases}
\end{eqnarray}
where $J_m,H^{(1)}_m$ are Bessel and Hankel functions of the order $m$.\cite{Olver10}

The boundary conditions are
\begin{equation}\label{bound-condition}
\mathcal{H}_{in}\Big|_{\gamma} = \mathcal{H}_{out}\Big|_{\gamma}\;,\quad
\frac1{\eps}\frac{\partial\mathcal{H}_{in}}{\partial r}\Big|_{\gamma} = \frac{\partial\mathcal{H}_{out}}{\partial
    r}\Big|_{\gamma}\;,
\end{equation}
where $\gamma$ is the contour separating $\mathcal{D}^-$ and $\mathcal{D}^+$ domains. Applying Green's theorem \cite{Jackson} to Eq. (\ref{Green-integral2D}) we can reduce the surface integral as
\begin{eqnarray}\label{Green-integral1D}
\mathcal{H}_{in}(\vec{r})=\int_{\gamma}\vec{n}\Big[\mathcal{H}_{out}(\vec{r}')\nabla G_p(\vec{r},\vec{r'})
-\eps G_p(\vec{r},\vec{r'})\nabla\mathcal{H}_{out}(\vec{r}')\Big]dl',\nonumber\\
\mathcal{H}_{out}(\vec{r})=-\int_{\gamma}\vec{n}\Big[\mathcal{H}_{in}(\vec{r}')\nabla G(\vec{r},\vec{r'}) -
\frac{1}{\eps}G(\vec{r},\vec{r'})\nabla\mathcal{H}_{in}(\vec{r}')\Big]dl'+\nonumber\\
+\int_{\Gamma}\vec{n}\Big[\mathcal{H}_{out}(\vec{r}')\nabla G(\vec{r},\vec{r'}) -
G(\vec{r},\vec{r'})\nabla\mathcal{H}_{out}(\vec{r}')\Big]dl'\;,
\label{integral}
\end{eqnarray}
where $dl'$ is the element of path, $\vec{n}$ is the unit vector along the external normal, $\Gamma$ is some remote contour (Fig.~\ref{fig:infty}). The integral over $\Gamma$ in the last line can be calculated explicitly by the known relation for the Wronskian determinant: \cite{Olver10}
\begin{eqnarray*}
    -\frac{k\rho}{4i}\int\limits_{-\pi}^\pi\left[{H}_0^{(1)}(k\rho)+i\cos\varphi {H}_1^{(1)}(k\rho)\right]e^{ik\rho\cos\varphi}\,d\varphi=\\=2\pi k\rho
    \left[J_1(k\rho){H}_0^{(1)}(k\rho)-J_0(k\rho){H}_1^{(1)}(k\rho)\right]=1.
\end{eqnarray*}
Then this integral  reproduces the field of a plane incident wave,
$\mathcal{H}^{(0)}_{out}=\mathcal{H}_0e^{i\vec{k}\vec{r}}$. Eqs. (\ref{integral}) are similar to boundary integral equations; the only difference is the absence of the factor 1/2 in the terms outside the integral.
These terms are given within the external or internal limit, in contrast to boundary equations, where they are determined directly at the contour. \cite{Kern:09}

The successive approximation series can be built up for both external and internal fields:
\begin{equation}\label{Expansion}
\mathcal{H}_{out} = \mathcal{H}^{(0)}_{out} + \mathcal{H}^{(1)}_{out} +
\dots,\quad \mathcal{H}_{in} =
\mathcal{H}^{(0)}_{in} + \mathcal{H}^{(1)}_{in} + \dots
\end{equation}
Then from (\ref{Green-integral1D})
we get the recurrent relations:
\begin{eqnarray}\label{Green-integral-recursion}
\mathcal{H}^{(j)}_{in}(\vec{r})=\int_{\gamma}\vec{n}\Big[\mathcal{H}^{(j)}_{out}(\vec{r}')\nabla
G_p(\vec{r},\vec{r'}) -\eps G_p(\vec{r},\vec{r'})
\nabla\mathcal{H}^{(j)}_{out}(\vec{r}')\Big]dl',\nonumber\\
\mathcal{H}^{(j+1)}_{out}(\vec{r})=-\int_{\gamma}\vec{n}\Big[\mathcal{H}^{(j)}_{in}(\vec{r}')\nabla
G(\vec{r},\vec{r'}) - \frac{1}{\eps}G(\vec{r},\vec{r'})\nabla\mathcal{H}^{(j)}_{in}(\vec{r}')\Big]dl'\;.
\end{eqnarray}
The approximation exactly takes into account the boundary conditions that is distinguished from the Born approach in quantum mechanics. It is to emphasize, that the shape of the contour $\gamma$ can be arbitrary;
the circular cylinder (considered in the next section) is, basically, just the simplest example. The dielectric region $\mathcal{D}^-$ could be inconnected; in that case the contour $\gamma$ is a sum of all the boundaries of dielectric domains.

\section{Scattering by two cylinders}\label{sec:pair}
\begin{figure}[htbp]
\begin{tikzpicture}[scale=2.5]
    \draw[blue] (-2,0) circle (3pt);
    \filldraw[blue] (-2,0) circle (1pt);
    \draw[blue] (-2.2,0.5) node {\large$\vec{E}_0$};
    \draw[blue] (-2.,-0.3) node {\large$\vec{H}_0$};
    \draw[thick,->] (-2.,0)--(-1.5,0);
    \draw (-1.75,-0.15) node {\large$\vec{k}$};
    \draw[->,blue,thick] (-2,0)--(-2,0.5);
    \draw[red,line width=2pt] (-.25,0) circle (1);
    \draw[thick,->] (-0.25,0)--(1,0);
    \draw (1,-0.25) node {\large$x$};
    \draw[thick,->] (-0.25,0)--(-0.25,1.25);
    \draw (-0.5,1.25) node {\large$y$};
    \draw (-0.65,-.5) node {\large$\eps,a$};
    \draw[blue,line width=2pt] (2,1) circle (0.7);
    \draw[thick,->] (2,1)--(3,1);
    \draw (3,0.75) node {\large$\Tilde{x}$};
    \draw[thick,->] (2,1)--(2,2);
    \draw (1.75,2) node {\large$\Tilde{y}$};
    \draw (2,.5) node {\large$\eps_p,b$};
    \draw[very thick,->] (-0.25,0)--(2,1);
    \draw (1,.75) node {\large$\vec{R}$};
    \draw[very thick,->] (-0.25,0)--(1.25,1.5);
    \draw (0.65,1.05) node {\large$\vec{r}$};
    \draw[very thick,->] (2,1)--(1.25,1.5);
    \draw (1.75,1.35) node {\large$\Tilde{\vec{r}}$};
    \draw[very thick,->] (-0.25,0)--(1.25,-0.5);
    \draw (0.9,-0.6) node {\large$\vec{r}'$};
    \draw[very thick,->] (2,1)--(1.25,-0.5);
    \draw (1.7,0.) node {\large$\Tilde{\vec{r}}'$};
\draw[green!80!black,thick] (0.35,0) arc (0:33:0.4cm);
\draw (0.5,0.15) node {\large$\alpha$};
\end{tikzpicture}
\caption{The scheme of $p$-wave scattering by two parallel cylinders. The dielectric permittivity and the radius are indicated in the first (left) and second (right) cylinders. External infinitely remote  contour
$\Gamma$ is not shown.} \label{f2}
\end{figure}
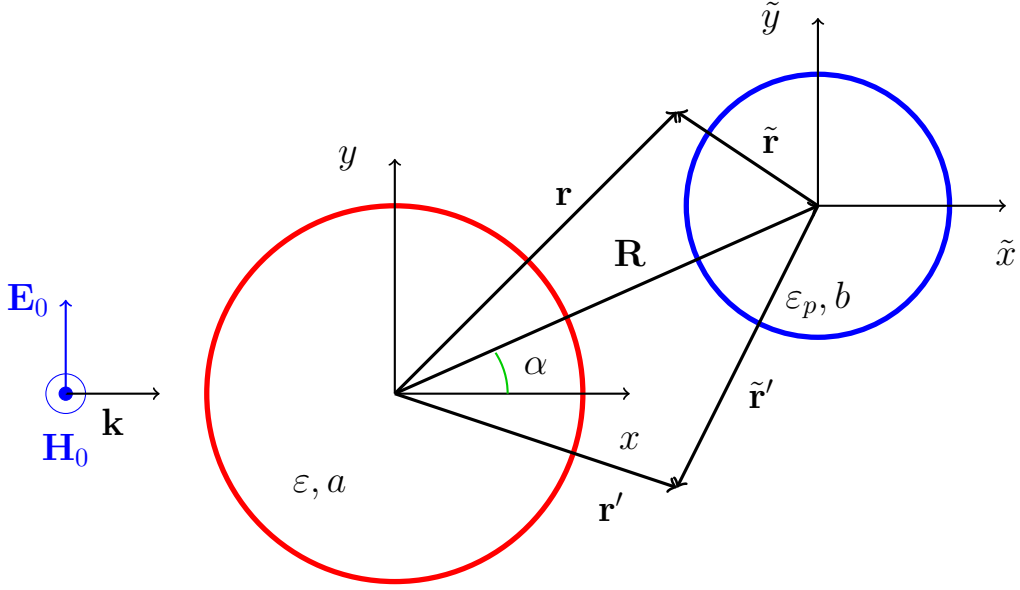

Let us now consider two cylinders, see Fig.~\ref{f2}.
There are three domains with different dielectric permittivity. The Helmholtz equation (\ref{Helmholtz-equation}) is valid for $k_0=\omega/c$, $k=\sqrt{\eps}\omega/c$ or
$k_p=\sqrt{\eps_p}\omega/c$, and the boundary conditions (\ref{bound-condition}) at the contour $\gamma\cup\gamma_p$ is:
\begin{equation}\label{Conditions}
\mathcal{H}_{in}\Big|_{\gamma} = \mathcal{H}_{out}\Big|_{\gamma},\;
\frac{1}{\eps}\frac{\partial\mathcal{H}_{in}}{\partial r}\Big|_{\gamma} = \frac{\partial\mathcal{H}_{out}}{\partial
    r}\Big|_{\gamma}\;,\quad
\mathcal{H}_{p}\Big|_{\gamma_p} = \mathcal{H}_{out}\Big|_{\gamma_p},\;
\frac{1}{\eps_p}\frac{\partial\mathcal{H}_{p}}{\partial \Tilde{r}}\Big|_{\gamma_p} = \frac{\partial\mathcal{H}_{out}}{\partial
    \Tilde{r}}\Big|_{\gamma_p}.
\end{equation}

Here, we treat the second cylinder as the perturbation. Let us obtain a number of successive approximations for the whole complicated configuration, shown in Fig.~\ref{f2}. We exploit the Green function for cylindric geometry given by Eq. (\ref{partial}). Using this Green function makes it possible to account for the first cylinder exactly including the multiple scattering. The second cylinder is described approximately in terms of the Born series. To found the number of terms, that would be sufficient to get the field with given accuracy, we compare it with a known well-studied numerical solutions such as discrete dipole approximation (DDA) and boundary element methods (BEM). 

The coupled boundary integral equations are analogous to Eqs. (\ref{Green-integral1D}). While the perturbation remains weak, the expansion (\ref{Expansion}) yields
\begin{eqnarray}
\mathcal{H}^{(j)}_{p}(\vec{\Tilde{r}})=%
\int_{\gamma_p}\vec{n}
\Big[\mathcal{H}^{(j)}_{out}(\vec{r}')
\nabla{G}_{p}(\vec{\Tilde{r}},\vec{\Tilde{r}'}) -\eps_p {G}_{p}(\vec{\Tilde{r}},\vec{\Tilde{r}'})
\nabla\mathcal{H}^{(j)}_{out}(\vec{r}')\Big]d\Tilde{l}',\nonumber\\
\mathcal{H}^{(j+1)}_{out}(\vec{r})=
-\int_{\gamma_p}\vec{n}\Big[\mathcal{H}^{(j)}_{p}(\vec{\Tilde{r}}')
\nabla G(\vec{r},\vec{r'}) - \frac{1}{\eps_p}G(\vec{r},\vec{r'})
\nabla\mathcal{H}^{(j)}_{p}(\vec{\Tilde{r}}')\Big]d\tilde{l}',\label{Integral_recursion}
\end{eqnarray}
where $\vec{n}=\vec{n}_{\gamma_p}$, $\Tilde{\vec{r}}=\vec{r}-\vec{R}$.
The recurrence relations (\ref{Integral_recursion}) are valid for arbitrary shape of the perturber with a sharp boundary, provided its layout is in the external region of the main cylinder. Further generalization for arbitrary shape of the first cylinder requires other Green function.

The integral over boundary of perturber can be calculated. The final relation is a series with a shift due to the axes offset:
 \begin{eqnarray}
 \mathcal{H}_p^{(j)}(\vec{\Tilde{r}})=\sum_{m=-\infty}^{\infty}e^{im\Tilde{\varphi}}J_m(k_p\Tilde{r})D^{m(j)}_{p},\nonumber\\
 \mathcal{H}_{out}^{(j)}(\vec{r})=\sum_{m=-\infty}^{\infty}e^{im\varphi}H_m(k_0r)D^{m(j)}_{out}+\sum_{m=-\infty}^{\infty}e^{im\Tilde{\varphi}}H_m(k_0\Tilde{r})\Tilde{D}^{m(j)}_{out}.
 \end{eqnarray}
 Coefficients $D$ are given by relations:
    \begin{eqnarray}
D^{m(0)}_{p}=\frac{\pi k_pb}{2i}
\Big[J_m(k_0b){H'}_m(k_pb)-\sqrt{\eps_p}{J'}_m(k_0b)H_m(k_pb)\Big]\times\nonumber\\
\times\Big[i^me^{ik_0R\cos\alpha}+\sum_{n=-\infty}^{\infty}i^ne^{i(n-m)\alpha}C_nH_{n-m}(k_0R)\Big],\nonumber\\
D^{m(j)}_{p}=\frac{\pi k_pb}{2i}\Bigl(\Tilde{D}^{m(j)}_{out}
\Big[H_m(k_0b){H'}_m(k_pb)-\sqrt{\eps_p}{H'}_m(k_0b)H_m(k_pb)\Big]+\nonumber\\
+\sum_{n=-\infty}^{\infty}D^{n(j)}_{out}e^{i(n-m)\alpha}H_{n-m}(k_0R)\Big[J_m(k_0b){H'}_m(k_pb)-\sqrt{\eps_p}{J'}_m(k_0b)H_m(k_pb)\Big]\Bigr);\nonumber\\
D^{m(j+1)}_{out}=-\frac{\pi k_0b}{2i}C_m\sum_{n=-\infty}^{\infty}D_{p}^{n(j)}e^{i(n-m)\alpha}H_{m-n}(k_0R)\Big[J_n(k_pb){J'}_n(k_0b)-\frac1{\sqrt{\eps_p}}{J'}_n(k_pb)J_n(k_0b)\Big],\nonumber\\
\Tilde{D}^{m(j+1)}_{out}=-\frac{\pi k_0b}{2i}D_{p}^{m(j)}\Big[J_m(k_pb){J'}_m(k_0b)-\frac1{\sqrt{\eps_p}}{J'}_m(k_pb)J_m(k_0b)\Big],\nonumber
\end{eqnarray}
where coefficients $C_m$ are given by Eq. (\ref{coeff}).

\begin{figure}
    \center{\includegraphics[width=0.6\linewidth]{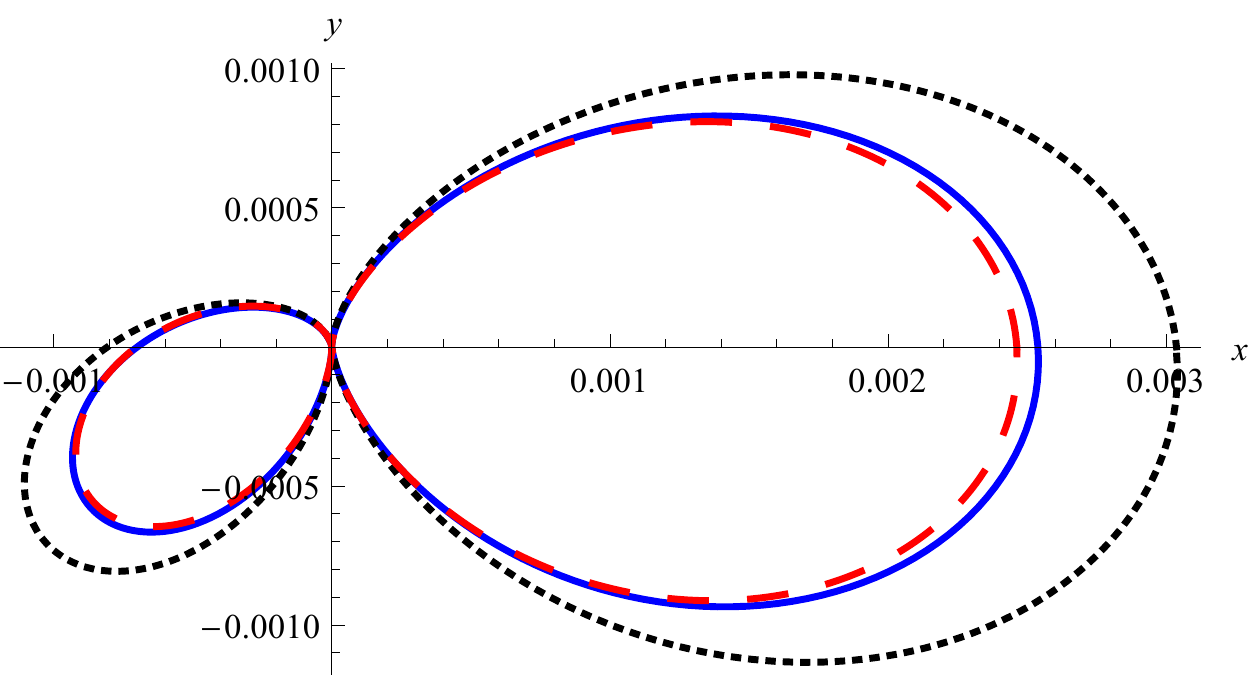}}
    \caption{Polar diagram of scattering by a pair of equal dielectric cylinders at $a=b=0.1\ \mu$m, $\eps_p=\eps=2.25$, $R=0.3~\mu$m at incidence angle $\alpha=-\pi/4$, the wavelength  $\lambda=1.5\ \mu$m, the distance between observation point and center of first cylinder is $r=2\lambda$: first Born approximation (dotted), second (dashed), and BEM (solid  line).}
    \label{fig:orders}
 \end{figure}

\begin{figure}
    \center{\includegraphics[width=0.6\linewidth]{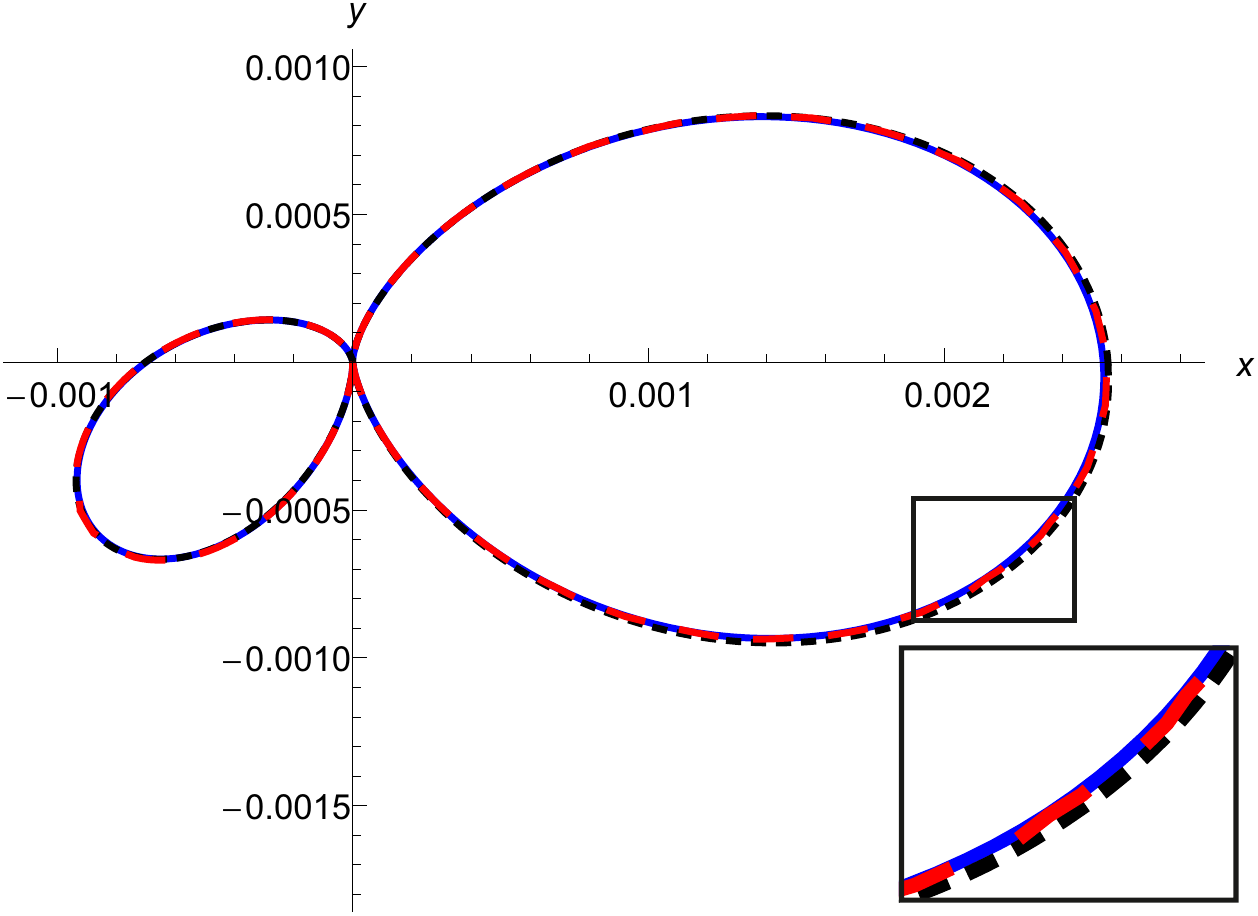}}
    \caption{Polar diagram of the scattered field at the same parameters as in Fig.~\ref{fig:orders}: the 3-rd Born approximation (dotted), BEM (solid), DDA (dashed line).
    The inset is the magnified part of the main plot, indicated by a square.
    }
        \label{fig:precision}
\end{figure}

Fig.~\ref{fig:orders} shows the angular dependence of scattered field square $|H_{sc}|^2$. As the figure demonstrates, the first approximation gives rather correct qualitative description of the diagram with a
deviation of 15\%. The error of the second order is nearly 3\%.
Fig.~\ref{fig:precision} shows the comparison of 3-rd Born approximation with numerical calculations by BEM and DDA. The deviation for 3-rd order appears to be about 1\%.

\section{Conclusions}
The Green function for a dielectric cylinder is found in the cases of $p$- and $s$-wave with source points inside and outside the cylinder. High-order Born approximation of two dielectrics with step boundaries are reduced to recurrence relations. This technique is analytically applied to the scattering by a pair of cylinders. The first approximation demonstrates its qualitative agreement in shape with numerical results. The second and third approximations are shown to agree quantitatively with calculation by boundary elements and discrete dipoles.

\appendix
\section{Scalar Green function}\label{sec:Green}

\begin{figure}[htbp]\centering
    \begin{tikzpicture}[scale=1.5]
    \draw[fill=gray!40] (0,0) circle (1cm);
    \draw[thick,->,blue, line width=2pt] (0,0)--(0,2);
    \draw[thick,->,blue, line width=2pt] (0,0)--(2,0);
    \filldraw[color=green!70!black] (1.06,1.06) circle (.1cm);
    \draw (2.2,0) node {\large$x$};
    \draw (0,2.2) node {\large$y$};
    \filldraw[blue] (-3,0) circle (2pt);
    \draw[blue] (-3,0) circle (5pt);
    \draw[blue] (-3.5,1) node {\large$\vec{E}_0$};
    \draw[blue] (-3,-0.6) node {\large$\vec{H}_0$};
    \draw[thick,->] (-3,0)--(-1.5,0);
    \draw (-2,-0.25) node {\large$\vec{k}$};
    \draw (1.25,0.75) node {\large$(x,y)$};
    \draw[->,blue,thick] (-3,0)--(-3,1);
    \end{tikzpicture}
    \caption{The $xy$-plane cross section of an infinite cylinder in homogeneous space. A circle indicates the observation point $\vec{r}=(x,y)$. The polarization of $p$-wave is shown at the left.} \label{f1}
\end{figure}
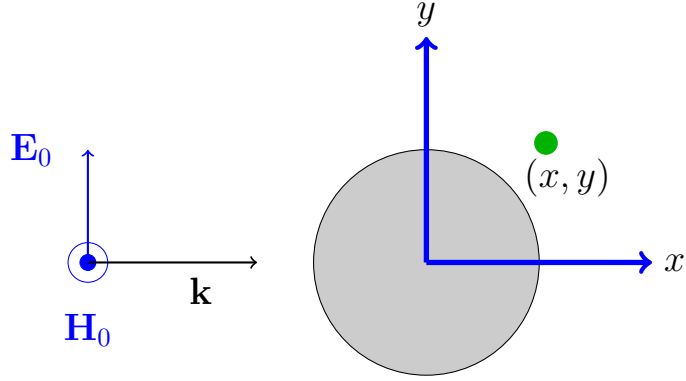

Let us consider a cylinder, which axis is along $z$ direction, as shown in Fig.~\ref{f1}.
We are looking for scalar Green function
$G(\mathbf{r}, \mathbf{r}')$ that is the solution to the
inhomogeneous 2-dimensional Helmholtz equation (\ref{Helmholtz-equation}) with $k=k_0$ in free space and $k=k_1$ in the dielectric.

Rewrite delta function (\ref{Helmholtz-equation}) in polar coordinates
\begin{equation}\label{polar}
\delta(\vec{r}-\vec{r}\,')=\frac1r\delta(r-r')\delta(\varphi-\varphi'),
\end{equation}
where $r,\varphi$ и  $r',\varphi'$  are the polar coordinates of source and observation points and decompose
the angular factor into the Fourier series:
\begin{equation}\label{Fourier}
\delta(\varphi-\varphi')=\frac1{2\pi}\sum_{m=-\infty}^\infty e^{im(\varphi-\varphi')}.
\end{equation}
The coefficient $1/2\pi$ is found from the delta-function normalization $\int_{-\pi}^{\pi}\delta(\varphi)\,d\varphi=1.$

Expanding the Green function in partial waves
\begin{equation}\label{Separation}
G(r,\varphi;r',\varphi')=\sum_{m=-\infty}^\infty g_m(r,r') e^{im(\varphi-\varphi')}
\end{equation}
and substituting into (\ref{Helmholtz-equation}) we get an ordinary equation for each $m$:
\begin{equation}\label{Differential}
\frac{d^2g_m}{dr^2}+\frac1r\frac{dg_m}{dr}+\left(k^2-\frac{m^2}{r^2}\right)g_m=\frac1{2\pi r}\delta(r-r').
\end{equation}
At $r>r'$ or $r<r'$, the corresponding solutions can be expressed through the combinations of Bessel  and Hankel functions:
\begin{equation}\label{cases}
g_m(r,r')=
\begin{cases}
a_mJ_m(k_1r),&r<a,\\
A_mJ_m(kr)+B_mH^{(1)}_m(kr),&a<r<r',\\
d_mH^{(1)}_m(kr),&r'<r.
\end{cases}
\end{equation}

Conditions (\ref{bound-condition}) are the continuity of the magnetic field's and its weighted normal derivative at the interface between dielectric and free space, $r=a$:
\begin{equation}\label{Maxwell-boundary}
\left[g_m\right]_{r=a}=0,\quad \left[\frac1{\eps}\frac{dg_m}{dr}\right]_{r=a}=0,
\end{equation}
where the square bracket denotes a jump of the corresponding value. The conditions are written for $p$-wave, where the magnetic field is parallel to the $z$-axis. Next pair of conditions follow from the continuity
of Green function and the jump of its first derivative at $r=r'$:
\begin{equation}\label{jump}
\left[g_m\right]_{r=r'}=0,\quad \left[\frac{dg_m}{dr}\right]_{r=r'}=\frac1{2\pi r'}.
\end{equation}
We omit Hankel function in the first line and Bessel function in the third line in Eq. (\ref{cases}) on a basis of regularity at $r\to0$ and Sommerfeld radiation requirement at $r\to\infty$.

Substituting (\ref{cases}) into boundary conditions (\ref{Maxwell-boundary}), (\ref{jump}) we get the set for the coefficients:
\begin{eqnarray}
a_mJ_m(k_1a)=A_mJ_m(ka)+B_mH^{(1)}_m(ka),\nonumber\\
\frac1{\sqrt\eps}a_mJ'_m(k_1a)
=A_mJ'_m(ka)+B_mH^{(1)\prime}_m(ka),\label{dielectric}\\
A_mJ_m(kr')+B_mH^{(1)}_m(kr')-d_mH_m^{(1)}(kr')=0,\nonumber\\ A_mJ'_m(kr')+B_mH^{(1)\prime}_m(kr')
-d_mH_m^{(1)\prime}(kr')=-\frac1{2\pi kr'}.\label{source}
\end{eqnarray}
Here the prime means the derivative of cylindric functions with respect to their arguments.

From (\ref{dielectric}) we get $B_m=\alpha_mA_m$, where
\begin{equation}
\alpha_m=\frac{J_m(ka)J'_m(k_1a)-\sqrt{\eps}J'_m(ka)J_m(k_1a)}
{\sqrt{\eps}H^{(1)\prime}_m(ka)J_m(k_1a)-H^{(1)}_m(ka)J'_m(k_1a)}.
\label{alpha}
\end{equation}
Then the determinant of the set (\ref{source}) for coefficients $A_m,d_m$ is $ \{J_m+\alpha_mH^{(1)}_m,H_m^{(1)}\}={2i}/{\pi kr'}, $ where the curly bracket stands for the Wronskian determinant $\{f,g\}=fg'-f'g$ at
$r=r'$. The final form of (\ref{cases}) is
\begin{eqnarray}\label{partial}
g_m(r,r')=
\begin{cases}
\frac1{4i}\left[J_m(ka)+\alpha_mH^{(1)}_m(ka)\right]H^{(1)}_m(kr')\frac{J_m(k_1r)}{J_m(k_1a)},&r<a,\\
\frac1{4i}\left[J_m(kr)+\alpha_mH^{(1)}_m(kr)\right]H^{(1)}_m(kr'),&a<r<r',\\
\frac1{4i}\left[J_m(kr')+\alpha_mH^{(1)}_m(kr')\right]H^{(1)}_m(kr),&r'<r.
\end{cases}
\end{eqnarray}

The Green function for $s$-wave can be obtained in a similar way, replacing the boundary conditions by the continuity of function $g_m$ together with its first derivative at $r=a$ instead of Eq.
(\ref{Maxwell-boundary}). Besides, the results would differ when the source is outside the dielectric.

Let us summarize the formulas for partial Green function. At {$r'< a$} they are:
\begin{eqnarray}
g_m(r,r')=\begin{cases}
\frac1{4i}J_m(k_1r\phantom{'})\big(C_mJ_m(k_1r')+H_m(k_1r')\big),&\;\;0\phantom{'}< r\phantom{'} < r',\\
\frac1{4i}J_m(k_1r')\big(C_mJ_m(k_1r\phantom{'})+H_m(k_1r\phantom{'})\big),&\;\;r'< r\phantom{'} < a,\\
\beta_mH_m(k_0r)J_m(k_1r'),&\;a\phantom{'} < r;\\
\end{cases}\label{hidden}
\end{eqnarray}
\begin{eqnarray}
C_m=-\frac{H_m(k_1a)H_m'(k_0a)
    -\eps^\nu H_m'(k_1a)H_m(k_0a)}{\Delta},\quad
\beta_m=\frac{\eps^\nu}{2\pi k_1a\Delta};\nonumber
\end{eqnarray}
At {$r'>a$} the formulas are:
\begin{eqnarray}
g_m(r,r')=
\begin{cases}
\beta_mJ_m(k_1r\phantom{'})H_m(k_0r'),&\;\;r\phantom{'} < a,\\
\frac1{4i}H_m(k_0r')\big(J_m(k_0r\phantom{'})+C_mH_m(k_0r\phantom{'})\big),&\;\;a\phantom{'} < r\phantom{'} < r',\\
\frac1{4i}H_m(k_0r\phantom{'})\big(J_m(k_0r')+C_mH_m(k_0r')\big),&\;\;a\phantom{'} < r' < r;\\
\end{cases}\label{visible}
\end{eqnarray}
\begin{eqnarray}
C_m=
-\frac{J_m(k_1a)J_m'(k_0a)
    -\eps^\nu J_m'(k_1a)J_m(k_0a)}{\Delta},\quad
\beta_m=\frac{1}{2\pi k_0a\Delta};\label{coeff}\\
\Delta=J_m(k_1a)H_m'(k_0a)
-\eps^\nu J_m'(k_1a)H_m(k_0a).\nonumber
\end{eqnarray}
Here the upper index $(1)$ of Hankel function is omitted. The formulas with $\nu=-1/2,1/2$ refer to the case of $p$ or $s$ wave, respectively. Expressions for $p$- or $s$-wave differ in the factor
${\eps}^{-1/2}$ or ${\eps}^{1/2}$ due to distinct boundary conditions. The Eq. (\ref{partial}) reduces to the particular case of (\ref{visible}) with $\nu=-1/2.$

\section{DDA}\label{sec:DDA}

Below we briefly recall 2-dimensional DDA approach\cite{Martin98} to obtain here the particular relationships we used in our calculations. Let us have some scattering body, with the volume $V$ (which is per unit
length along $z$ direction in 2-dimensional case) and the permittivity $\varepsilon$ (which is constant within the body), placed in vacuum. From the Helmholtz equation we obtain the integral equation for isotropic
medium:
\begin{equation}
\label{Integral_general}
\mathbf{E}(\mathbf{r})=\mathbf{E}_{inc}(\mathbf{r})+\int_{V\backslash
V_0}d^2 r'\left[\widehat{\mathbf{G}}(\mathbf{r},\mathbf{r}')
\chi(\mathbf{r}')\mathbf{E}(\mathbf{r}')\right] +\int_{V_0}d^2
r'\left[\widehat{\mathbf{G}}(\mathbf{r},\mathbf{r}')
\chi(\mathbf{r}')\mathbf{E}(\mathbf{r}')\right],
\end{equation}
where $V_0$  is a small  volume around singularity point $\vec{R}=\vec{r}-\vec{r'}\to0$, $V\backslash V_0$ is the volume of dielectric without the singular part, $\mathbf{E}_{inc}(\mathbf{r})$ is the given field of incident wave, $\chi(\mathbf{r}) \equiv (\varepsilon - 1)/4\pi$ is the polarizability, the Green tensor
$\widehat{\mathbf{G}}(\mathbf{r},\mathbf{r}')$ is the solution to Maxwell equations:
\begin{equation}
\label{Green tensor_def} {\rm rot}\, {\rm rot}\,
\widehat{\mathbf{G}} - k^2 \widehat{\mathbf{G}} = 4 \pi k^2 \delta
(\mathbf{r} - \mathbf{r}').
\end{equation}

The Green tensor obeying (\ref{Green tensor_def}) can be expressed \cite{Novotny_book,Kern:09} in terms of scalar Green function $g$ that satisfies Eq. (\ref{Helmholtz-equation})
\begin{equation}
\label{Green tensor_vs_scalar} G_{\alpha\beta} =
4\pi\left(k^2\delta_{\alpha\beta} + \nabla_\alpha
\nabla_\beta\right) g,
\end{equation}
where $\alpha$, $\beta$ --- are Cartesian indices. Then, it is
well known that the Green tensor actually depends on the
difference $\mathbf{R}$. Finally, in 2-dimensional case we have
\begin{eqnarray}
\label{G_final} G_{\alpha\beta} (\mathbf{R}) = \frac{i \pi k}{R}
\left[A(k R)\delta_{\alpha\beta} - B(k R) \frac{R_\alpha
R_\beta}{R^2}\right], \\ \nonumber A (x) = x H_0^{(1)}(x) -
H_1^{(1)}(x), \quad \nonumber B (x) = x H_0^{(1)}(x) - 2H_1^{(1)}(x),
\end{eqnarray}
where $H_0^{(1)}(x)$, $H_1^{(1)}(x)$ are Hankel functions of the first kind.

In (\ref{Integral_general}) we implicitly isolate the term, that
includes the singularity of the Green tensor at $\mathbf{r} =
\mathbf{r}'$, by means of a small volume $V_0$, for which the
point $\mathbf{r}$ is internal. Then, we rewrite this term,
introducing the following quantities:
\begin{equation}
\label{M} \widehat{\mathbf{M}}(V_0, \mathbf{r}) = \int_{V_0}d^2 r'
\left[G_{\alpha\beta} (\mathbf{r}-\mathbf{r}') - \frac{4 R_\alpha
R_\beta - 2\delta_{\alpha\beta}
R^2}{R^4}\right]\chi(\mathbf{r}')\mathbf{E}(\mathbf{r}')
\end{equation}
and
\begin{equation}
\label{L} \widehat{\mathbf{L}}(V_0, \mathbf{r}) = - \int_{V_0}d^2
r' \frac{4 R_\alpha R_\beta - 2\delta_{\alpha\beta}
R^2}{R^4}\chi(\mathbf{r}')\mathbf{E}(\mathbf{r}'), \qquad
\mathbf{R}\equiv \mathbf{r} - \mathbf{r}'.
\end{equation}
Note that $\widehat{\mathbf{M}}$ is free from the singularity,
thus $\widehat{\mathbf{M}}\rightarrow 0$ with $V_0 \rightarrow 0$.
The fraction under integration is, basically, the static limit (at
$k \rightarrow 0$) of the Green tensor. Also, we need to
discretize the whole scattering volume $V$ into the parts $V_j$
(in such a way that $V_0$ coincides with one of them). With the
use of (\ref{M}) and (\ref{L}) the equation
(\ref{Integral_general}) becomes:
\begin{equation}
\label{Integral_sampled}
\mathbf{E}(\mathbf{r}_i)=\mathbf{E}_{inc}(\mathbf{r}_i)+\sum_{j\ne
i}\int_{V_j}d^2
r'\left[\widehat{\mathbf{G}}(\mathbf{r}_i-\mathbf{r}')
\chi(\mathbf{r}')\mathbf{E}(\mathbf{r}')\right]+\widehat{\mathbf{M}}(V_i,
\mathbf{r}_i) - \widehat{\mathbf{L}}(V_i, \mathbf{r}_i),
\end{equation}
where $\mathbf{r}_i$ denotes a point lying inside the volume
$V_i$.

Up to this line, the equations are fully correct as being exact
consequences of the initial wave equation. Now
we make two approximations: the first is that
$\mathbf{E}(\mathbf{r}')$ and $\chi(\mathbf{r}')$ are constant
within the volume $V_j$; the second approximation assumes that
\begin{equation}
\label{Gij} \frac{1}{V_j} \int_{V_j}d^2 r'
\widehat{\mathbf{G}}(\mathbf{r}_i,\mathbf{r}') =
\widehat{\mathbf{G}}(\mathbf{r}_i,\mathbf{r}_j).
\end{equation}
The condition (\ref{Gij}) is
intrinsically contained in all DDA formulations,\cite{Yurkin07} which initially
deal with replacing the scatterer with a set of point dipoles. If
the volumes $V_i$ are square cells (we should keep in mind that we
are treating 2-dimensional case) then we can place the points
$\mathbf{r}_i$ to the center of the corresponding squares.

Below, we will neglect $\widehat{\mathbf{M}}$, as most of authors
do, choosing by that the simpler (or "weak")
DDA formulation.\cite{Lakhtakia92,Yurkin07}
Integrating (\ref{L}) we transform (\ref{Integral_sampled})
into its final form
\begin{equation}
\label{DDA_dipoles} \mathbf{d}_i \widehat{\mathbf{\alpha}}_i^{-1}
= \mathbf{E}_{i, inc} + \sum_{j\ne
i}\widehat{\mathbf{G}}(\mathbf{r}_i - \mathbf{r}_j) \mathbf{d}_j,
\end{equation}
where we denote, for simplicity, the dependence on $\mathbf{r}_i$
(and $\mathbf{r}_j$) by the corresponding subscript; $\mathbf{d}_i
= V_i\chi_i \mathbf{E}_i$ --- the polarization of the volume $V_i$
(basically, its dipole moment, as we took $\chi_i$ and
$\mathbf{E}_i$ being constant within $V_i$); and
$\widehat{\mathbf{\alpha}}_i$ is the polarizability tensor defined
as
\begin{equation}
\label{DDA_alpha} \widehat{\mathbf{\alpha}}_i =
\widehat{\mathbf{I}} V_i\chi_i\left(1 + 2\pi \chi_i\right)^{-1}
\equiv \frac{a^2}{2}\frac{\varepsilon - 1}{\varepsilon + 1}
\widehat{\mathbf{I}}.
\end{equation}
The last term is the known quasi-static dipole polarizability of a
cylinder (2-dimensional dipole) with the cross section, $\pi a^2$,
equal to $V_i$.

Thus, the calculations consisted in finding the dipole moments
$\mathbf{d}_i$ by solving (\ref{DDA_dipoles}) with
(\ref{DDA_alpha}) and (\ref{G_final}). Upon them, all the
quantities of interest can be obtained. In our case, we calculate
the scattered magnetic field.

\section*{Acknowledgements}

Authors are grateful to O. V. Belai for helpful discussions. This work is supported by the Russian Foundation of Basic Research \# 16-02-00511 and the Government program of the leading research schools NSh-6898.2016.2.

%

\end{document}